# Silicon-based high-temperature diamondlike ferromagnetic with self-organized superlattice distribution of manganese impurity

E.S. Demidov, E.D. Pavlova, A.I. Bobrov

*Nizhni Novgorod State University, Gagarin avenue 23, Nizhni Novgorod 603950, Russia*

Crystal structure of Si:Mn diluted magnetic semiconductor with Curie temperature 500K, deposited from laser plasma on GaAs(100) substrates, was investigated by high-resolution transmission electronic microscopy and diffraction. This laser technology allows to reach of solid solution of 15%Mn in silicon with high electrical and full magnetic activity, conservation of diamondlike crystal structure and epitaxy growth of Si:Mn. The self-organized formation of superlattice structures takes place with period equal to trebled distance between nearest atomic layers (110) and interval between layers (110) which are doped by Mn atoms and oriented along direction of growth of 50 nm Si:Mn film.

To necessity to fill a gap between microelectronics and magnetic electronics J. K. Furdyna[1] had paid attention in 1988. The first ferromagnetic so-called diluted magnetic semiconductors (DMS) on basis of compounds II-VI[2,3] and III-V[4] were synthesized in 90s of last century. Now there are a lot of publications devoted to epitaxy layers of DMS $Ga_{1-x}Mn_xAs$ with atomic fraction of Mn impurity x≈0.05 and perfect crystal structure grown up at low temperatures ≈250°C by molecular beam epitaxy (MBE). However maximum attained Curie point of this ferromagnetic $Ga_{1-x}Mn_xAs$ 170K is below room temperature[5]. Especially interesting DMS are on basis of elementary silicon semiconductor because of their compatibility with the most widespread silicon technology. Numerous attempts of synthesis ferromagnetic DMS on basis of silicon, doped by 3d – impurities, were undertaken[6-22]. The problem of synthesis such DMS is even more difficult than in case of III-V compounds because of the close to equilibrium solubility of middle 3d - iron group elements impurities in Si is about $10^{16}$ cm$^{-3}$ on two order more low, these impurities mainly occupy position of interstitials and show donor properties[23,24]. Though, according to regularities in recharge levels of such impurities in diamond like semiconductors, the most preferable analogue of manganese in compounds III-V in silicon is iron[18,25], the greatest successes have been reached by doping of Si also by manganese. Encouraging preconditions have appeared still in 1993[6]. The ferromagnetic behavior up to 320K according to field dependence of magnetization was observed for periodic structure of 20 nm layers of Si and Mn fabricated by electron beam vacuum evaporation (EBVE). The layers of DMS Si:Mn with about 5% Mn and $T_c$=70K, according to abnormal Hall effect (AHE), were produced on silicon substrates (Si:Mn/Si) at 300°C by MBE technology as well as in case of DMS on basis of compounds III-V[7]. The single crystal layers of Si:5%Mn/Si were made by EBVE method[9,10] with ferromagnetism on magnetization data up to 400K. However DMS's rather high resistivity (0.25-2.5) Ohm·cm with semiconductor dependence $\rho=\rho_0\exp(E_a/kT)$ means absence of degeneration of semiconductor. The degeneration is necessary for realization of the most probable in DMS ferromagnetic spin ordering of 3d - ions by RKKY indirect exchange interaction through current carriers – ho les[18,26]. Hence for p <$10^{20}$ cm$^{-3}$ ferromagnetism[9-11] is caused by certain inclusions of second phases. The method of pulsed deposition from laser plasma (PLD) was applied by authors[13] for synthesis at 400-700°C of DMS Si:(2-10)%Mn with appearance of ferromagnetism up to 400K which was related to $Mn_5Si_2$, $Mn_{81.5}Si_{18.5}$ and $Mn_5Si_3$ inclusions. For the first time the electron diffraction perfect single crystal layers of Si:(2-5)%Mn were formed by implantation of Mn$^+$ ions with energy 200 keV and subsequent 5 min annealing at 600-900°C. However value $T_c$=70K was small, p <1.2·$10^{18}$ cm$^{-3}$, the ferromagnetism was more likely caused by non-uniform distribution of manganese. Authors[15,16] fabricated high $T_c$ Si:Mn DMS layers with application of Mn$^+$ ions implantation in Si (300 keV[15], 195 keV[16]) and subsequent 5-10min annealing at 800°C. In paper[15] for layers with 0.8% peak atomic fraction of Mn the ferromagnetism, according to magnetization, was shown up to 400K. In work[16] the layers with concentration of Mn (2.7-5.9)·$10^{20}$ cm$^{-3}$ showed ferromagnetism up to 305K on magnetization and Faraday effect data.

The most high-temperature DMS on basis of diamondlike semiconductors GaSb, InSb, Ge and Si doped by Mn or Fe were synthesized in our laboratory by PLD [17-18]. The mainly investigated by us DMS Si:Mn with 10-15 % Mn has greatest mobility of current carriers, the Mn impurity in it shows high electric and magnetic activity. The ferromagnetism of DMSs was confirmed by observation of ferromagnetic resonance (FMR), AHE and magnetooptic Kerr effect (MOKE). The fact that DMS Si:Mn has some ordered structure but ferromagnetic behavior not been caused by any inclusions was checked up by experiments with an overheat, ionic irradiation, comparison with discrete alloys with the same components. The usual reflective x-ray and electron diffraction do not allowe to define this structure (in x-ray diffraction because of small thickness of layers and too small weight of Si atoms and in electron diffraction possibly because of masking by amorphous surface oxide layer). Here we represent new data about structure of our layers DMS Si:Mn/GaAs recently obtained with application of high resolution transmission electronic microscopy (HRTEM) and selected area electron diffraction (SAED).

The PLD technology fabrication of DMS Si:Mn on single crystal GaAs substrates with (100) orientation was described in [17,18]. Measurements of cross-section HRTEM and SAED were made on JEOL device JEM-2100F. Investigated Si:Mn layers by thickness of 50 nm were grown at 300°C in which, as well as in[17], atomic fraction of manganese on x-ray spectral analysis data was 15%. The MOKE, AHE, high hole conductivity with resistivity $\rho$=2.5·$10^{-4}$ Ohm·cm and mobility of holes $\mu$=33 cm$^2$/V·s were registered. The concentration of holes p ≥ 7.5·$10^{20}$cm$^{-3}$ means that not less than 10% of Mn impurity atoms is electric active. As well as in[17-19], at room temperature FMR spectra consisted of several peaks of resonant absorption which with decreasing of temperature merge in single rather narrow peak. According to FMR data at 93K, if to consider that manganese spin is equal 5/2, the concentration $N_{Mn}$=8·$10^{21}$cm$^{-3}$ practically coincides with x-ray value 15%Mn ($N_{Mn}$=7.5·10$^{21}$cm$^{-3}$), i.e. manganese is totally magnetic active.

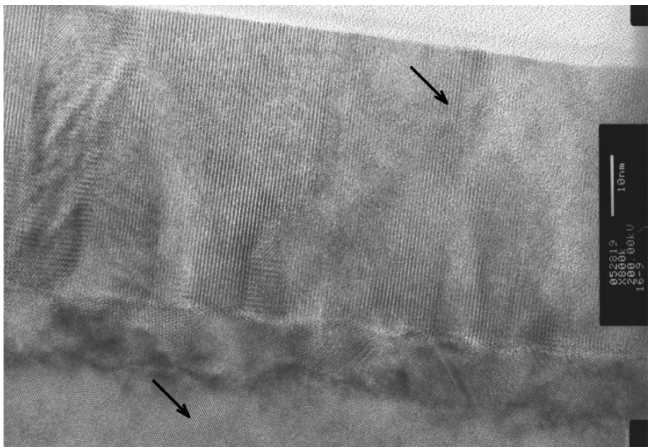

**Fig. 1.** Cross-section HRTEM lattice image along <110> direction of PLD deposited at 300°C 50nm layer Si:15%Mn on single crystal substrate GaAs with (100) orientation.

The cross-section HRTEM lattice image along direction <110> of deposited from laser plasma at 300°C 50 nm layer Si:15%Mn on single crystal substrate GaAs with orientation (100) is shown on Fig.1. Apparently the film Si:Mn, except for block distortions of crystal structure, is mainly epitaxial with diamondlike structure in which the lattice planes (110) are oriented perpendicularly to border of Si:Mn/GaAs and are parallel to similar (110) planes of GaAs substrate.

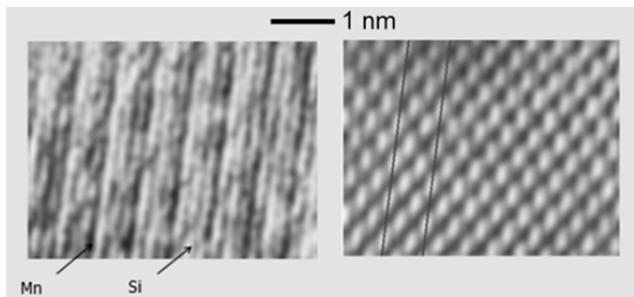

**Fig. 2.** Scaled up images marked by arrows of small sites on Fig. 1, Si:Mn layer on the left and GaAs substrate on the right. On left designated by arrows dark points correspond to probable localization of impurity Mn atoms on places of Si atoms in (110) planes and light points correspond to Si atoms. On right for ruled visualization two parallel dark lines allocate GaAs (110) substrates planes with distance between them equal distance at the left between partially replaced by Mn atoms (110) layers in the film Si:Mn.

Fig. 2 shows the scaled up images marked by arrows of small sites on Fig. 1 Si:Mn layer and GaAs substrate. At the left designated by arrows dark points correspond to probable localization of Mn impurity atoms on the places of Si atoms in (110) planes and light points correspond to Si atoms. The blacked out view of Mn against Si atoms at electron translucence of sample is caused by negative charge localized on Mn impurity acceptor centers according to recharge levels on Fig.1 in article[18] and introduced by this impurity high hole conductivity of deposited from laser plasma Si:Mn films[17-19]. Differing from silicon part of sample chess alternation of dark and light points on the right on fig.2 is obviously caused by raised negative charge on As atoms on comparison with Ga atoms in partially ionic III-V GaAs crystal substrate. On this part of Fig. 2 for ruled visualization two parallel dark lines allocate GaAs (110) planes with distance between them equal distance at the left between partially replaced by Mn atoms (110) layers in the film Si:Mn. The same distance is equal trebled space interval between crystal planes (110) as in film Si:Mn and GaAs substrate. That is the crystal periodicity of Si:15%Mn is close to that at GaAs and some more than in not doped silicon. The most interesting consists of that according to Fig. 2 in PLD synthesized DMS Si:15%Mn takes a place the superlattice self-organized distribution of manganese in planes (110) in its solid solution in silicon. These enriched by manganese (110) planes are oriented along the direction of growth of Si:Mn layer and perpendicularly boundary between this layer and substrate GaAs probably because of temperature gradient taking a place during laser deposition. That fact that in every third atomic (110) layer of Si atoms with manganese partially replaced atoms of silicon means that concentration of this impurity is less than 30% in the consent with 15% Mn on the x-ray analysis data. Other important fact consists that according Fig. 1 there are no obvious inclusions of a second phase in accordance with uniform distribution of magnetization on picture of MFM cross-section chipping in[18] for similar sample Si:Mn/GaAs and single line of FMR at 93K. The splitting of FMR spectrum on some peaks of absorption with temperature rise (see fig. 1 in[19]) is caused by presence of block distortions of crystal structure of Si:Mn layer instead of non-uniform distribution of Mn impurity in the DMS as it was supposed[19].

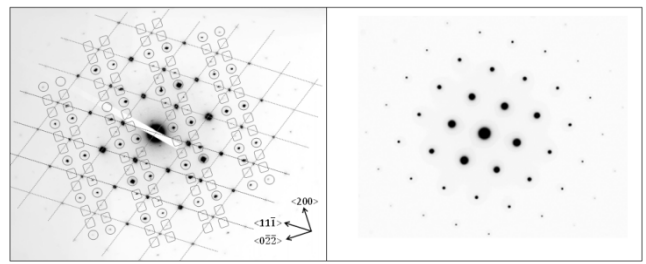

**Fig. 3.** SAED patterns of Si:Mn layers at the left and substrates GaAs on the right along the direction <110>. The crossings of thin lines at the left allocate the same reflexes of diamondlike lattice as for the single crystal substrate GaAs on the right. The circles and rounded squares allocate additional reflexes. Each series of the equidistant circles being on one line along a direction $<0\overline{22}>$ allocated with circles and crossings of thin lines of reflexes shows presence of superlattice modulation of diamond crystal structure with the trebled period along a $<0\overline{22}>$ direction perpendicularly crystal planes of (110) type.

The patterns of SAED along the direction <110> of Si:Mn layer and GaAs substrate are shown on Fig. 3. Crossings of thin lines in electronic diffraction of Si:Mn allocate the same reflexes of a diamondlike lattice as for single crystal GaAs substrate. Presence of such reflexes in addition to cross-section of Si:Mn layer in direct space on Fig. 2 proves diamondlike structure investigated Si:Mn DMS. The distance between these reflexes practically coincides with those for GaAs. It also in addition to Fig. 2 shows that the crystal periodicity of the Si:15%Mn DMS is close to that for GaAs and more than for undoped silicon.

The circles and rounded squares allocate additional reflexes. Each series of the equidistant circles being on one line along a direction $<0\bar{2}\bar{2}>$ allocated with circles and crossings of thin lines of reflexes shows in consent with HRTEM on Fig. 2 presence of superlattice modulation of diamond crystal structure with trebled period along a $<0\bar{2}\bar{2}>$ direction perpendicularly crystal planes of (110) type. Additional reflexes allocated with rounded squares from Si:Mn on Fig. 3 possibly are linked with presence of visible on Fig.1 of block distortions of crystal structure of it DMS layer.

The shown by HRTEM and SAED on Fig. 1-3 periodic sublattice self-organizing of solid solution of Mn impurity in the Si:15%Mn layers synthesized by laser method obviously should lead to essential difference of electronic band structures of it DMS in comparison with undoped silicon. There is probably higher than shown on the energy diagram of fig. 1 in article [18] position of the valence band top in relation to the acceptor recharge levels of substituting Si by Mn atoms in the Mn doped silicon. It provides experimentally observed according to HE and FMR the high electric and magnetic activity of Mn impurity in investigated layers of Si:Mn. The demonstrated possibility of epitaxy growth of high-temperature silicon ferromagnetic with diamondlike structure on a single crystal substrate of not magnetic diamondlike semiconductor is important for applications. At the same time, this unique ferromagnetic represents the big academic interest as for physics of superfast formation of the supersaturated by transition elements of single crystal solid solutions with periodic subnanosized self-organizing and for development of electronic band theory of solutions of spin polarized 3d - ions with the superlattice structure.

Authors are grateful prof. Pavlov D.A. for the help in carrying out of measurements and useful discussion and Pitirimova E.A. for useful discussion. Supported by RFBR (08-02-01222-a, 11-02-00855-a), the Ministry of Education of Russian Federation (projects 2.1.1/2833 and 2.1.1/12029) and the State contract № 02.740.11.0672 of the Federal purpose program «Scientific and scientific-pedagogical cadre of innovative Russia» 2009-2013 is acknowledged.

[1] J. K. Furdyna, **J. Appl. Phys. 64**, R29 (1988).
[2] H. Ohno, H. Munekata, T. Penney, S. von Molna´r, and L. L. Chang, **Phys. Rev. Lett. 68**, 2664 (1992).
[3] H. Ohno, A. Shen, F. Matsukura, A. Oiwa, A. Endo, S. Katsumoto, and Y. Iye, **Appl. Phys. Lett. 69**, 363 (1996).
[4] H. Munekata, H. Ohno, S. von Molnár, A. Segmüller, L. L. Chang, and L. Esaki, **Phys. Rev. Lett. 63**, 1849 (1989).
[5] Maciej Sawicki, in Spintronic Materials and Technology, Series in Materials Science and Engineering, ed. By Y.B. Xu, S.M. Thompson CRC Press Taylor & Francis Group, 2007, P. 57-76.
[6] T. Takeuchi, Y. Hirayama, M. Futamoto, **IEEE Trans. Magn. 29**, 3090 (1993).
[7] H. Nakayama, H. Ohta, T. Kulatov, **Physica B. 302-303**., 419 (2001).
[8] H.M Kim, N.M Kim, C.S.Park, S.U.Yuldashev, T.W.Kang, K.S. Chung, **Chem. Mater. 15**, 3964 (2003).
[9] F.M. Zhang, Y. Zeng, J. Gao, X.C. Liu, X.S. Wu, Y.W. Du, **JMMM 282**, 216 (2004).
[10] H.K. Kim, D. Kwon, J.H. Kim, Y.E. Ihm, D. Kim, H. Kim, J.S. Baek, C.S. Kim, W.K. Choo, **JMMM 282**, 244 (2004).
[11] D. Kwon, H.K. Kim, J.H. Kim, Y.E. Ihm, D. Kim, H. Kim, J.S. Baek, C.S.Kim, W.K. Choo, **JMMM 282**, 240 (2004).
[12] F.M. Zhang, X.C. Liu, J. Gao, X.S. Wu, Y.W. Du, **Appl. Phys. Lett. 85**, 786 (2004).
[13] Sug Woo Jung, Gyu-Chul Yi, Yunki Kim, Sunglae Cho, and J. F. Webb, **Electronic Materials Letters 1**, 53 (2005).
[14] Y.H. Kwon, T.W. Kang, H.Y. Cho, T.W. Kim, **Sol. St. Comm. 136**, 257 (2005).
[15] M. Bolduc, C. Awo-Affouda, A. Stollenwerk, M.B. Huang, F.G. Ramos, G. Agnello, V.P. LaBella, **Phys. Rev. B 71**, 033302(2005).
[16] A.B. Granovskii, Yu.P. Sukchorukov, A.F. Orlov, N.S. Perov, A.V. Korolev, E.A. Gan'shina, V.I. Zinenko, Yu.A. Agafonov, V.V. Saraikin, A.V. Telegin, D.G. Yarkin, **JETP Letters 85**, 335 (2007).
[17] E.S. Demidov, Yu. A. Danilova, V.V. Podolskii, V.P. Lesnikov, M. V. Sapozhnikov, A. I. Suchkov, **JETP Letters 83**, 568 (2006).
[18] E.S. Demidov, V.V. Podolskii, V.P. Lesnikov M. V. Sapozhnikov, D. M. Druzhnov, S. N. Gusev, B. A. Gribkov, D. O. Filatov, Yu. S. Stepanova, and S. A. Levchuk, **JETP 106**, 110 (2008).
[19] E. S. Demidov, B. A. Aronzon, S. N. Gusev, V. V. Karzanov, A. S. Lagutin, V. P. Lesnikov, S. A. Levchuk, S. N. Nikolaev, N. S. Perov, V. V. Podolskii, V.V. Rylkov, M. V. Sapozhnikov, A.V. Lashkul, **JMMM 321**, 690 (2009).
[20] Shin Yabuuchi1, Yukinori Ono, Masao Nagase, Hiroyuki Kageshima, Akira Fujiwara, and Eiji Ohta, **Jap. Journ. Appl. Phys. 47**, 4487 (2008).
[21] Li Zeng, E. Helgren, M. Rahimi, and F. Hellman, R. Islam and B. J. Wilkens, R. J. Culbertson and David J. Smith, **Phys. Rev. B 77**, 073306 (2008).
[22] T.T. Lan Anh, S.S.Yu, Y.E. Ihm, _, D.J. Kim, H.J. Kim a, S.K. Hong, C.S. Kim, **Physica B 404**, 1686 (2009).
[23] G. W. Ludwig and H. H. Woodbury, *Electron Spin Resonance in Semiconductors,* Wiley, New York, 1962.
[24] E. R. Weber and N. Wiehl, in *Proceedings of Symposium on Defects Semiconductors 2,* Boston, Mass., 1982, p. 19.
[25] E. S. Demidov, **Fiz. Tverd. Tela (St. Petersburg) 34**, 37 (1992) [Sov. Phys. Solid State **34**, 18 (1992)].
[26] T. Dietl, H. Ohno, F. Matsukura, **Phys. Rev. B 63,** 195205 (2001).